\documentclass[conference]{IEEEtran}

\usepackage{xcolor}
\usepackage{tikz}

\usepackage{float,graphicx}

\usepackage{hyperref}
\usepackage{cleveref}
\usepackage{xurl}

\definecolor{lime}{HTML}{A6CE39}
\DeclareRobustCommand{\orcidicon}{%
	\begin{tikzpicture}
		\draw[lime, fill=lime] (0,0)
		circle [radius=0.16]
		node[white] {{\fontfamily{qag}\selectfont \tiny ID}};
		\draw[white, fill=white] (-0.0625,0.095)
		circle [radius=0.007];
	\end{tikzpicture}
	\hspace{-2mm}
}
\newcommand{\orcidID}[1]{\href{https://orcid.org/#1}{\orcidicon}}

\usepackage{fullpage}
\usepackage[T1]{fontenc}

\begin{document}

\title{Post-Quantum Cryptography in the 5G Core}

\author{
	\IEEEauthorblockN{Thomas Attema\IEEEauthorrefmark{1}\IEEEauthorrefmark{2}\orcidID{0000-0002-8289-6853}, Bor de Kock\IEEEauthorrefmark{2}\orcidID{0000-0003-3143-4381}, Sandesh Manganahalli Jayaprakash\IEEEauthorrefmark{2}, \\ Dimitrios Schoinianakis\IEEEauthorrefmark{3}\orcidID{0000-0001-5228-6715}, Thom Sijpesteijn\IEEEauthorrefmark{2}, and Rintse van de Vlasakker\IEEEauthorrefmark{2}\thanks{Author list in alphabetical order.} \thanks{Document date: December 23, 2025.}}

\IEEEauthorblockA{\IEEEauthorrefmark{1}CWI, Amsterdam, The Netherlands}
\IEEEauthorblockA{\IEEEauthorrefmark{2}TNO, The Hague, The Netherlands}
\IEEEauthorblockA{\IEEEauthorrefmark{3}Nokia Bell Labs, Athens, Greece}
}


\IEEEoverridecommandlockouts

\maketitle

\newcommand{\TODO}[1]{\color{red}\sffamily \{~TODO: #1~\} \color{black}\normalfont}
\newcommand{\BOR}[1]{\color{blue}\sffamily \{~Bor: #1~\} \color{black}\normalfont}
\newcommand{\DIM}[1]{\color{magenta}\sffamily \{~Dimitris: #1~\} \color{black}\normalfont}
\newcommand{\INS}[1]{\color{magenta} #1 \color{black}\normalfont}
\newcommand{\DEL}[1]{\color{magenta} \st{#1} \color{black}\normalfont}

\begin{abstract}
    In this work, the conventional cryptographic algorithms used in the 5G Core are replaced with post-quantum alternatives and the practical impact of this transition is evaluated. Using a simulation environment, we model the registration and deregistration of varying numbers of user equipments (UEs) and measure the resulting effects on bandwidth consumption and latency.
    Our results show that the deployment of post-quantum cryptographic algorithms has a measurable effect on performance, but that this effect is small, and perhaps more crucially, that the extra overhead needed in terms of computation and bandwidth does not have any substantial impact on the usability of the network and the efficiency of its network functions. 
    Overall the experimental results in this work corroborate earlier research: the 5G Core is technically able to support post-quantum cryptography without any inherent issues connected to the increased computational overhead or larger message size. 
\end{abstract}
\section{Introduction}
Quantum computers will have a profound impact on the world as we know it: they will be able to perform computations that have so far been infeasible, thus realizing various breakthroughs in fields like chemistry and biology. 
Unfortunately, the emergence of quantum computing has introduced unprecedented risks to modern cryptographic frameworks --- the foundation of our digital security and economy --- necessitating urgent reconsideration of security protocols across digital infrastructures. Shor’s quantum algorithm threatens to compromise widely deployed public-key cryptographic systems by solving some mathematical problems, such as integer factorization and discrete logarithms, in polynomial time --- tasks that remain computationally infeasible for classical computers. As the mathematical foundation of many cryptographic algorithms is based on the assumption that these problems are impossible to solve, this is a problem ---  and this vulnerability extends to the cryptographic foundations of 5G networks, which rely heavily on these conventional protocols to secure data transmission, user authentication, and data integrity.

As 5G networks become the backbone of global connectivity, their long operational lifecycle raises concerns about future-proofing against quantum threats. Current encryption methods securing 5G communications will be broken once quantum computers achieve sufficient scale, exposing sensitive data to retroactive decryption attacks. This ``harvest now, decrypt later'' risk underscores the critical need to preemptively integrate quantum-resistant cryptographic primitives into the 5G Core (5GC) architecture.

By adopting standards for post-quantum cryptography (PQC), which rely on mathematical problems resistant to both conventional and quantum attacks, the 5G Core can mitigate vulnerabilities while maintaining compliance with evolving regulatory and industry requirements. Various standardization organizations (such as the IETF \cite{ietf-pquip-pqc-engineers-09}) and NIST in the United States \cite{nistNISTSpecial}) as well as policy makers (such as the European Commission \cite{europaRecommendationCoordinated}, BSI in Germany \cite{bundQuantumTechnologies} and NCSC in the United Kingdom \cite{thequantuminsiderSetsTimeline}) have started to encourage transitioning to post-quantum security within the next coming years --- or even require it in order for security certifications to be upheld. It is to be expected that mobile network technologies will have to undergo this transition as well, within the coming years.

\subsection{Background}

\paragraph{Post-Quantum Cryptography (PQC)} 
The development of large-scale quantum computers will have a large impact on the security of commonly used conventional cryptographic algorithms. Shor's algorithm from 1994 \cite{Shor94} can be used to mount devastating attacks on the public key algorithms that currently secure most internet communications, including RSA \cite{RivShaAdl78} and Diffie-Hellman \cite{DifHel76}. This means the development of new algorithms for key establishment and digital signatures is a requirement. To address this need, the United States National Institute of Standards and Technology (NIST) initiated a multi-year standardization effort to find new algorithms for key encapsulation (KEM) and digital signature algorithms (DSA), a process that at the time of writing has lead to its first published standards. This Post-Quantum Cryptography standardization process was highly transparent and in collaboration with the cryptographic research community at large \cite{SchwabePQC}. In contrast, the impact of quantum computing on symmetric cryptography is far more limited: although there has been a discussion on whether Grover's algorithm is able to brute-force 128-bit symmetric keys, NIST recently clarified that 128-bit keys for AES remain quantum-safe for decades to come \cite{nistAES}. 

In this work, our experiments rely on KEMs and digital signature algorithms from this NIST competition: for KEMs, we evaluate the code-based BIKE algorithm~\cite{NISTPQC-R4:BIKE22} as well as the lattice-based FrodoKEM~\cite{NISTPQC-R3:FrodoKEM20}. For digital signatures, our experiments include algorithms based on Module Lattices (ML), Stateless Hash functions (SLH), and Fast Fourier Transforms over NTRU lattices (FN). Specifically, ML-DSA is the standardized algorithm previously known as \emph{Dilithium}~\cite{NISTPQC:CRYSTALS-DILITHIUM22}, and SLH-DSA corresponds to the hash-based scheme formerly referred to as \emph{SPHINCS+}~\cite{NISTPQC:SPHINCS+22}. The lattice-based scheme \emph{Falcon}~\cite{NISTPQC:FALCON22} has also been selected for standardization, with its final standards document still under development~\cite{NIST_2024}; upon publication, \emph{Falcon} is expected to be renamed FN-DSA.
We also evaluate the performance of elliptic curve (\texttt{secp2561}) and RSA-signatures to, to compare their performance with their PQC equivalents.
For a thorough overview on post-quantum cryptography in general and the various `families' of algorithms, we refer the reader to e.g. \cite{bernstein2017post, EPRINT:Peikert14,EPRINT:Peikert15}. 

On a final note, we use the term \emph{conventional cryptography} to refer to cryptography from before the post-quantum transition, i.e. cryptography that was not built with resistance against quantum adversaries in mind. 

\paragraph{Hybridization of conventional and post-quantum cryptography}
\label{sec:hybridization}
The term \emph{hybrid cryptography} can refer to several concepts, but in this work we refer to the combination of a conventional and a post-quantum algorithm into one scheme that has (some of) the benefits of both. 
There are several reasons that justify the use of hybrid cryptographic schemes. First, at the beginning of the post-quantum standardization process by NIST, there was some uncertainty whether these new protocols would indeed turn out to be secure in the longer run; adding a layer of conventional cryptography as well, meant that the security of protocols would at the very least not decrease below the original level.

Another reason a combination of conventional and post-quantum can make sense is the attacker model for \emph{current} deployments of post-quantum cryptography. A common frame of reference are so-called \emph{harvest-now, decrypt-later} attacks, in which we assume that a nation-state adversary stores large amounts of encrypted data now, so that attacks can be mounted whenever quantum computers are indeed available to perform them. 
Under this assumption we require post-quantum secrecy, but not yet post-quantum authentication: if an attacker stores messages now they can later try and break the encryption, but will not achieve anything by breaking the authentication. 

A downside is that the combination of two schemes will always add computational overhead and can also increase the message size of the key --- although the latter can be mitigated by e.g. \textsf{xor}'ing the two protocols' keys instead of sending both \cite{EPRINT:CroPaqSte19}.

Using hybrid schemes is promoted by several standardization bodies,\cite{irtf-cfrg-hybrid-kems-03,etsi} and is natively supported in various applications including the TLS implementation we use in our experiments. Because of this --- and as we do not expect the added computational overhead to be problematic for our use case --- we opt to use hybridized versions of post-quantum algorithms instead of standalone post-quantum algorithms in this work. 

\paragraph{Transport-Layer Security (TLS)}
TLS is a security protocol used for a variety of internet applications, most visibly \texttt{https} web traffic. 
Due to its versatility in terms of modes and ciphersuites, variations of TLS are used in a range of protocols, including industrial applications and IoT home appliances.
Although variants using a pre-shared key exist, most commonly TLS starts with a handshake between client and server, where among other things the cryptographic primitives and key lengths are negotiated depending on the capabilities of both parties. Newer TLS versions offer high flexibility in terms of ciphersuites, making it easy to exchange insecure protocols for post-quantum protocols such as the ones mentioned in the previous paragraph.
In recent years much has been published about how to achieve full post-quantum security in TLS, either by replacing the classical signatures with a post-quantum alternative \cite{USENIX:BBCT22} or by changing the approach to authentication altogether. 
In KEMTLS, for example, signatures in the handshake are replaced with instantiations of KEMs from the NIST competition \cite{CCS:SchSteWig20}.
An overview of the various strategies to migrate TLS to a post-quantum version can be found in \cite{Wiggers24}.

\begin{figure*}[!t]
\centering
\includegraphics[width=\textwidth]{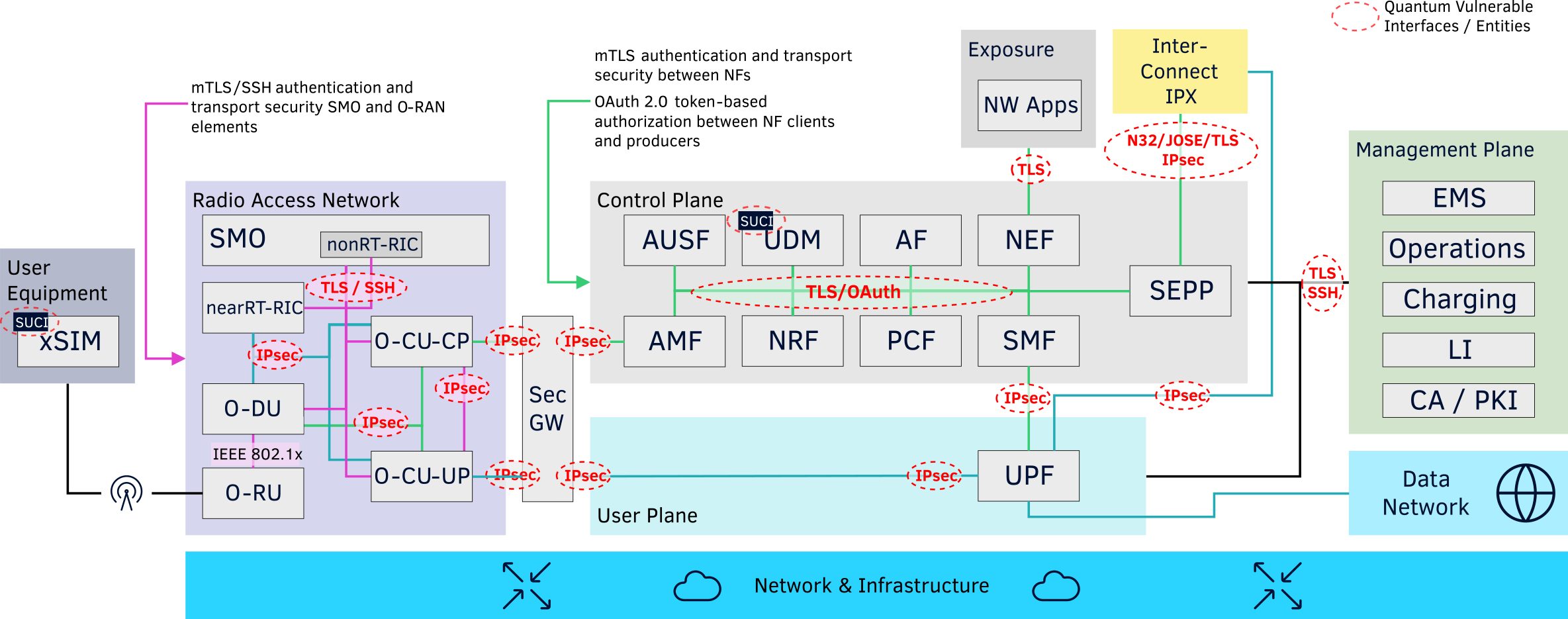}
\caption{Overview of vulnerable interfaces in a 5G network \cite{nbl}.} 
\label{fig:5G_PQC}
\end{figure*}

\paragraph{The 5G core network (5GC)}
The security architecture of 5G networks represents a marked departure from previous generations, reflecting fundamental changes in the design and deployment of authentication, authorization, and encryption mechanisms. Earlier generations of mobile networks relied predominantly on symmetric-key cryptography: the Subscriber Identity Module (SIM) stored a long-term shared secret that enabled mutual authentication between the user equipment (UE) and the network~\cite{Clancy2019PostquantumCA}. While effective in earlier deployment contexts, this model offered limited flexibility and did not readily accommodate the scale and heterogeneity of emerging services.

In contrast, 5G adopts a cloud-native, microservice-oriented core architecture and makes extensive use of Internet protocols and standards—most notably Transport Layer Security (TLS) and OAuth~2.0—to secure service-to-service communication~\cite{Clancy2019PostquantumCA}. This architectural transition is accompanied by a shift in the trust model: rather than relying primarily on long-term symmetric secrets, 5G core security increasingly depends on a Public Key Infrastructure (PKI)-based framework to support dynamic authentication and scalable key management~\cite{Clancy2019PostquantumCA}. Although this change improves operational agility, it also expands the attack surface with respect to quantum-capable adversaries, since public-key primitives are precisely the class of mechanisms threatened by Shor-type attacks~\cite{Clancy2019PostquantumCA,Scalise2024AnAA}.

More broadly, the 5G security architecture spans multiple domains, including access-network security, core-network security, and service-based architecture (SBA) security, each with distinct constraints that influence the integration of post-quantum cryptography (PQC)~\cite{Scalise2024AnAA,Hanna2024IntegratingPT}. The core network is of particular relevance in this context: its virtualized network functions (VNFs) and service-based interfaces both enable comparatively agile software updates and create practical opportunities for cryptographic migration~\cite{Scalise2024AnAA}. At the same time, virtualization and increased software complexity introduce additional risks (e.g., misconfiguration, expanded trust relationships, and interface exposure), underscoring the need to integrate PQC within a comprehensive and systematically engineered security framework~\cite{Scalise2024AnAA,Hanna2024IntegratingPT}.

\paragraph{Current Vulnerabilities and Security Requirements}
Despite its advanced security features, the 5G architecture contains inherent vulnerabilities that quantum computing may exploit. The transition to software-based network functions increases the attack surface and potential for exploitation through software vulnerabilities \cite{Scalise2024AnAA}. Additionally, the reliance on PKI for authentication and authorization makes 5G networks particularly susceptible to quantum attacks that can break the underlying mathematical problems of current public key cryptosystems \cite{Clancy2019PostquantumCA, Scalise2024AnAA}.

The security requirements for 5G networks are demanding and multifaceted, encompassing data confidentiality, integrity, authenticity, privacy, and availability \cite{Hanna2024IntegratingPT}. These requirements are further complicated by the diverse use cases 5G supports, from enhanced mobile broadband (eMBB) to ultra-reliable low-latency communications (URLLC) and massive Internet of Things deployments \cite{Scalise2024AnAA}. Each use case presents unique security challenges and performance constraints that must be considered when implementing post-quantum cryptography \cite{Ojetunde2024AMR}.

5G networks must also contend with the challenge of ensuring end-to-end security across a heterogeneous ecosystem of devices, networks, and service providers \cite{Uysal2023DataDrivenMD}. This complexity is amplified by the need to maintain interoperability with legacy systems while introducing quantum-resistant security measures \cite{Clancy2019PostquantumCA}. The security architecture must therefore be flexible enough to accommodate different security profiles and cryptographic capabilities while ensuring a consistent security posture across the network \cite{Hanna2024IntegratingPT, Ojetunde2024AMR}. 

In Figure \ref{fig:5G_PQC} a high-level overview of a 5G system is provided, highlighting the various interfaces and components that are vulnerable to quantum attacks. The figure emphasizes the need for a comprehensive security strategy that encompasses all aspects of the 5G architecture, from the radio access network (RAN) to the core network and service-based architecture \cite{nbl}. The RAN provides connectivity between user equipment and the core, carrying user traffic and enabling access to network services. The core network coordinates interactions among network functions and supports secure transport for both user-plane and control-plane communication. 

Figure \ref{fig:5G_PQC} also shows that asymmetric cryptography and in particular TLS is extensively used to protect all communication between different components of the core network. These components referred to as Network Functions are critical; they carry out all the control plane traffic of all the users on the network. Therefore, it is imperative to develop and share more implementations and performance benchmarks with the community to better understand the effects of post-quantum cryptography on these components. This paper aims to address this need by investigating the challenges associated with migrating the 5G core to post-quantum cryptographic solutions.

\subsection{Related work}
Several recent works focus on implementing and benchmarking post-quantum variants of the TLS protocol \cite{ESORICS:ZZDSZYZ24,sosnowskiPerformancePostQuantumTLS2023}, including optimizations for embedded devices \cite{ASIACCS:BKNS20,EPRINT:TDFZSS23}, and evaluations specifically targeted at the Android mobile operating system \cite{EPRINT:ManWigMoo23}. Notably, the work in \cite{sosnowskiPerformancePostQuantumTLS2023} reveals encouraging results for PQC adoption. For example, it was shown that HQC and Kyber perform comparably to legacy cryptography, while Dilithium and Falcon signature schemes demonstrate even faster performance. There were also no performance drawbacks observed from implementing hybrid cryptographic approaches. At higher NIST security levels, PQC algorithms actually outperformed currently deployed algorithms, although PQC might pose certain challenges when it comes to bandwidth-constrained applications \cite{sosnowskiPerformancePostQuantumTLS2023}.

Recent research on post-quantum cryptography (PQC) migration for 5G networks has focused on enhancing authentication protocols to address quantum threats. A notable contribution comes from researchers developing quantum-resistant extensions to the 5G Authentication and Key Agreement (AKA) protocol \cite{10.1145/3507657.3529657}.

There is still a considerable gap when it comes to the evaluation of PQC in actual telecom networks. This is extremely relevant considering also the fact that, the 3GPP, IETF, ITU and O-RAN standardization initiatives for the development of PQC telecommunication standards are still in progress, and so it is crucial to explore the migration challenges of PQC in real-world telecom environments.

\subsection{Our contributions}
In this work we replaced the key establishment and digital signature algorithms in 5G with post-quantum alternatives, and tested what impact this has on the network when deployed on a lab setup resembling a real-world 5G core network, where only the radio part of the network is simulated.

The work was done concurrently with \cite{Scalise2024AnAA} and is similar in idea and execution. Since the approach was slightly different we decided to make our experiment public. It should be seen as a confirmation of known/existing results rather than as a novel contribution. 

The main difference is that the authors of \cite{Scalise2024AnAA} implemented a custom open-source PQC-equipped \texttt{free5GCore} system to evaluate the initial handshake latency between VNFs. Their results suggest a negligible increase in UE connection setup duration and a small increase in connection setup data requirements. 
In contrast, in this work we considered a hybrid approach (described in Section \ref{sec:hybridization}), and included a migration of the signature algorithms to a post-quantum variant. We also provide a more thorough analysis of the impact on the network at varying numbers of devices. 
\section{Experiments}

\subsection{Set-up of the 5G core.} \label{subsubsec:enabling_pq_tls}
\begin{figure*}[ht]
    \begin{center}
        \includegraphics[width=.8\textwidth]{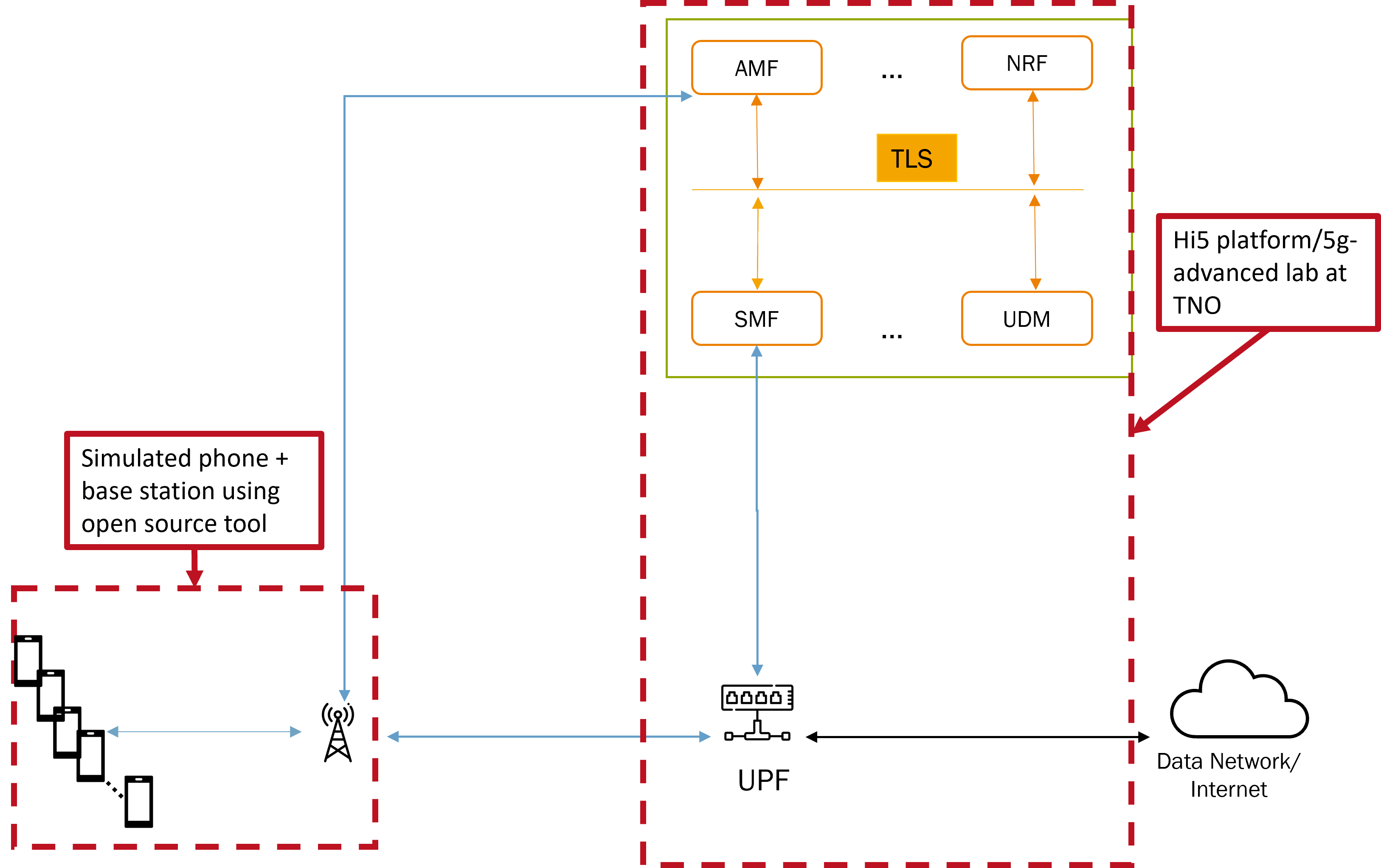}
        \caption{The setup for the experiments described in this work.}
        \label{fig:setup}
    \end{center}
\end{figure*}

The following experiments were performed in the 5G-lab at TNO, where we have access to an existing 5G core setup. 
An overview of the complete setup can be seen in Figure \ref{fig:setup}.
In our core setup we use an Intel Next Unit of Computing-machine with an i7-8559U processor (4 cores, hyperthreaded), to run a set of 10 network functions: \textsc{nrf, amf, upf, udr, udm, ausf, nssf, bsf, pcf,} and \textsc{smf}. When starting the core network, the required signature and KEM algorithms are configured on each of the network functions, and the necessary TLS certificates are generated and loaded depending on this specification. 

To make this possible, the following steps were first performed:
\begin{enumerate}
    \item \texttt{openssl} was compiled against
    \texttt{liboqs}~\cite{software:liboqs} to enable the \texttt{oqsprovider},
    granting access to PQ-TLS algorithms.
    \item \texttt{libcurl} was compiled against the \texttt{oqsprovider}-enabled
    \texttt{openssl}.
    \item \texttt{libngthttp2} was compiled against the \texttt{oqsprovider}-enabled
    \texttt{openssl}.
    \item \texttt{open5gs} was compiled against these custom versions of
    \texttt{libcurl} and \texttt{libngthttp2}.
    \item Changes were made to \texttt{open5gs}'s interface to \texttt{libcurl} and
    \texttt{libngthttp2} to be able to set KEM and signature algorithms as a
    parameter in the NFs.
    \item For each signature algorithm in the \texttt{oqsprovider}-enabled
    \texttt{openssl}, a set of PKI certificates was generated.
\end{enumerate}

Now we are able to start the 5G core with any selection of PQ-algorithms.

\subsection{Simulating radio connections.}
The user equipment (UE) and the Radio Access Network (RAN) were simulated using the open source tool \texttt{ueransim}~\cite{software:ueransim} on a separate Intel machine with the same specs as above. The two machines are connected over a 1 Gbit/s LAN connection over which the traffic between the radio access node and a core network is routed. \texttt{ueransim} enables us to create a large number of UEs using our experiment script. 

\begin{figure*}[ht]
\begin{center}
\includegraphics[width=.8\textwidth]{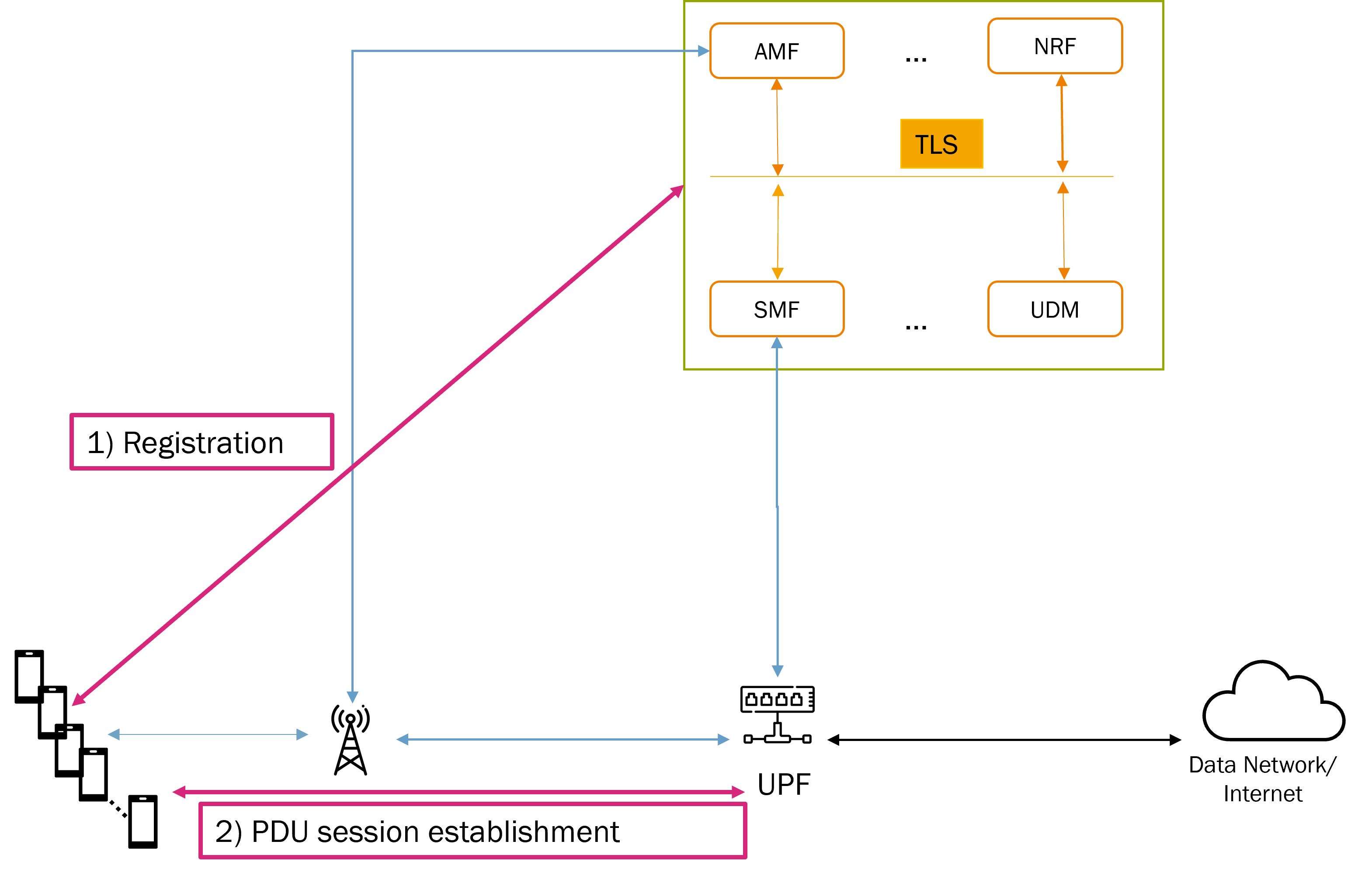}
\caption{A new registration request triggers setting up various connections in the core.}
\label{fig:registration}
\end{center}
\end{figure*}

At regular intervals, the simulated UEs send out a registration request (step (1) in Figure \ref{fig:registration}) followed by a PDU session creation request (step (2) in Figure \ref{fig:registration}) to the core network. For new UEs, this request then triggers setting up a secure connection between various network functions in the core network. 

\subsection{Choice of post-quantum algorithms.}
Table \ref{tab:algos} shows the algoritms chosen for our experiments. For BIKE, FrodoKEM, Falcon and ML-DSA a combination was made with a (conventional) ECC implementation, to achieve hybrid variants as described in Section \ref{sec:hybridization}.

\begin{table*}
    \begin{center}
        \caption{The Key Encapsulation Mechanisms and Digital Signature Algorithms used in the experiments. \\ Remark that \texttt{rsaEncryption} refers to RSA Signatures.}
        \label{tab:algos}
        \begin{tabular}{l l || l l }
            KEM & & DSA & \\
            \hline
            Plain ECC & \texttt{secp256r1} & RSA & \texttt{rsaEncryption} \\
            BIKE & \texttt{p256\_bikel1} & Falcon & \texttt{p256\_falcon513} \\
                 & \texttt{p384\_bikel3} & ML-DSA & \texttt{p384\_mldsa66}  \\
            FrodoKEM & \texttt{p521\_frodo1344shake} & SPHINCS+ & \texttt{sphincssha2129ssimple} 
        \end{tabular}
    \end{center}
\end{table*}

\subsection{Experimental setup.}
We run an experiment where we vary the number of active UEs. Each UE de-registers every $t$ seconds, after which it immediately re-registers and requests a PDU session. An attempt is made to space out the re-registrations in time by starting the initial registrations for each UE $t/n$ seconds apart, where $t$ is the re-registration interval in seconds (usually $t=10$) and $n$ is the number of UEs. 

\begin{figure*}[p]
    \begin{center}
        \includegraphics[width=\textwidth]{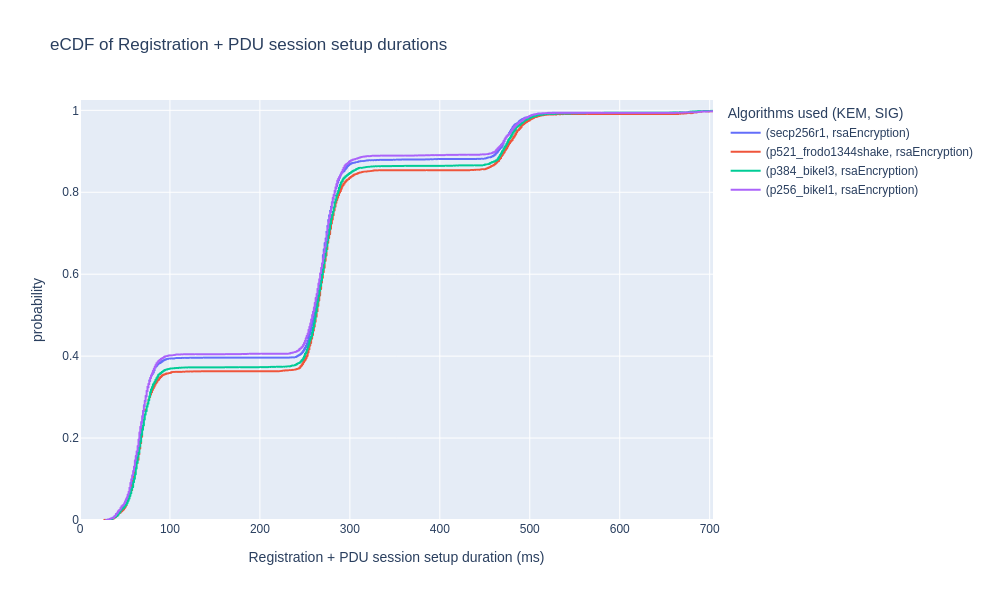}
        \caption{Histogram showing the setup duration with various KEMs. Remark that \texttt{rsaEncryption} refers to the RSA digital signature algorithm.}
        \label{fig:ecdf_pdusetup_kem}
    \end{center}
\end{figure*}

\begin{figure*}[p]
    \begin{center}
        \includegraphics[width=.9\textwidth]{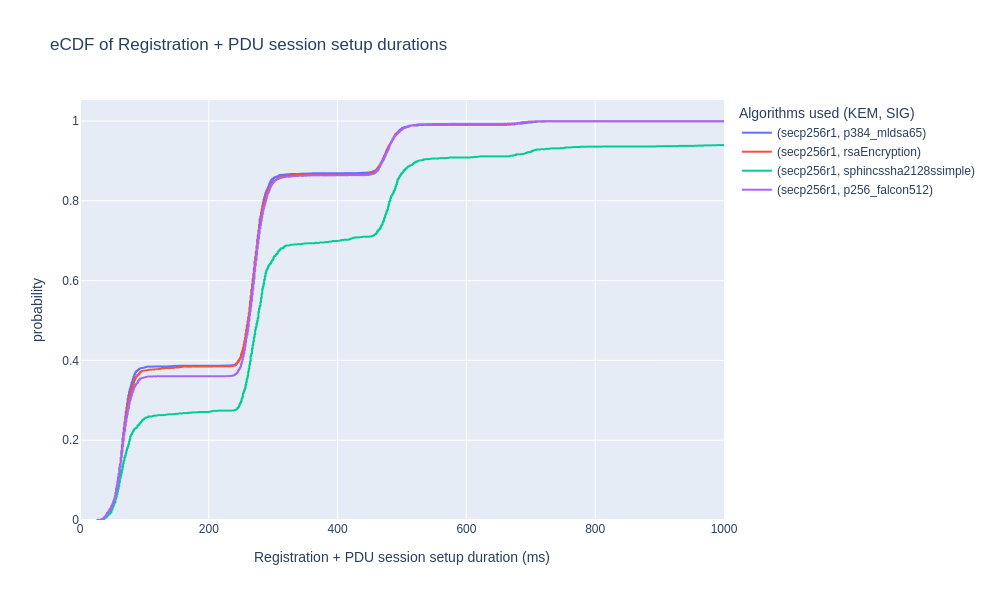}
        \caption{Histogram showing the setup duration with various signature algorithms, combined with a conventional (ECC-based) KEM.}
        \label{fig:ecdf_pdusetup_sig}
    \end{center}
\end{figure*}

\begin{figure*}[p]
    \begin{center}
        \includegraphics[width=.9\textwidth]{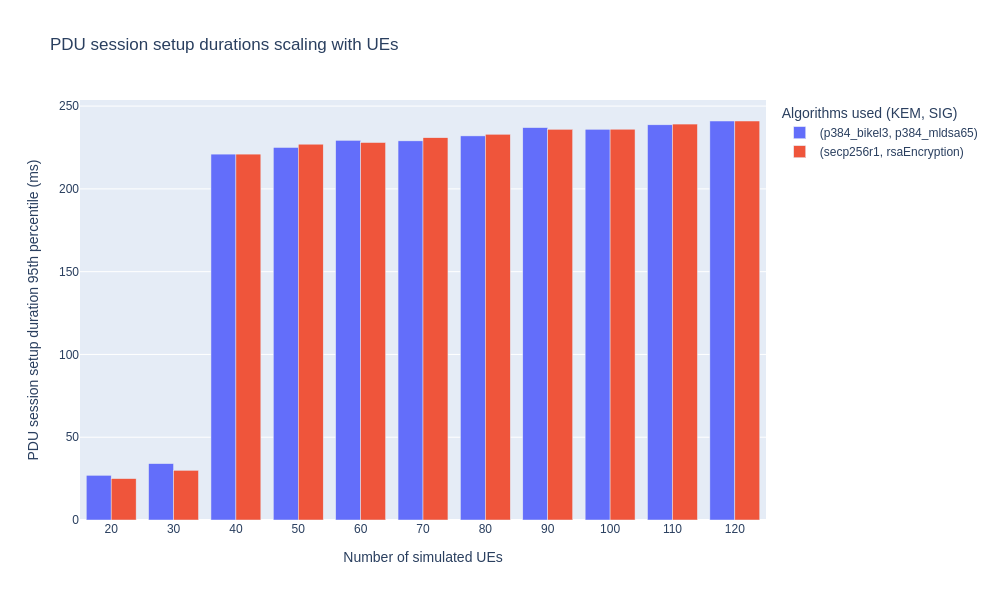}
        \caption{The duration of the slowest 95th percentile of registrations, for a varying number of UEs.}
        \label{fig:numUEscaling}
    \end{center}
\end{figure*}

\begin{figure*}[p]
    \begin{center}
        \includegraphics[width=\textwidth]{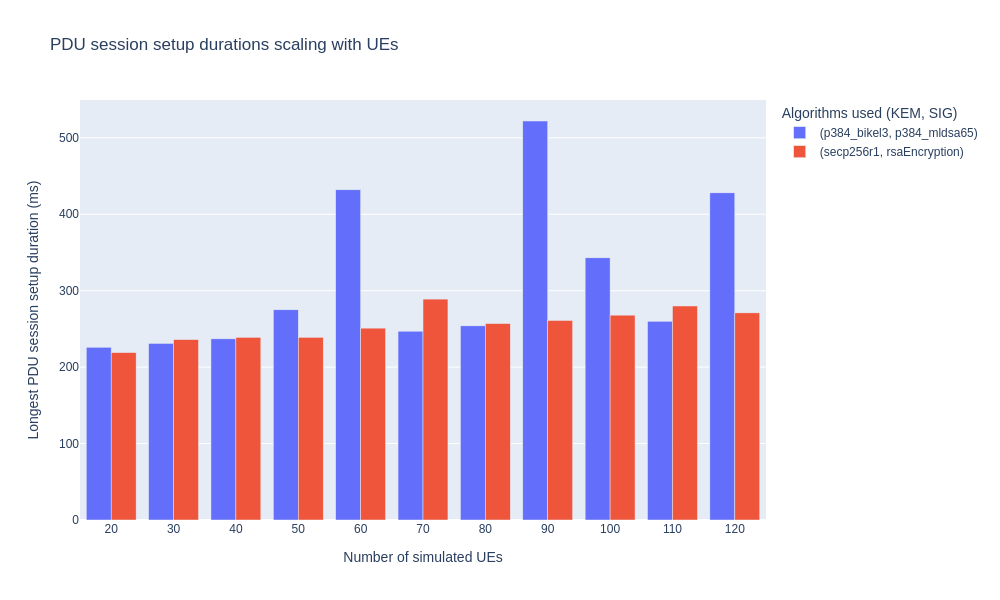}
        \caption{The duration of the slowest 100th percentile of registrations, for a varying number of UEs.}
        \label{fig:numUEscaling_100}
    \end{center}
\end{figure*}

The setup is built to measure both the latency and bandwidth impact of using these new algorithms, particularly focusing on the registration and PDU session establishment procedures. 
For latency the values indicate the total latency between the moment the \texttt{registration} request-message is sent and the moment the \texttt{registration accept-message} is received.
For bandwidth we measure the total amount of data in bytes sent from/to each network function's SBI interface. These measurements are performed using \texttt{bpftrace}. 

\subsection{Results.}
Both conventional and post-quantum algorithms are tested.
We first analyze the duration of device registration and session creation during setup for various algorithm choices. In Figure \ref{fig:ecdf_pdusetup_kem} we use RSA for signatures and vary the KEM used for each experiment. 
In Figure \ref{fig:ecdf_pdusetup_sig} we use conventional (elliptic curve-based) Diffie-Hellman for key agreement, but vary the signature scheme. 
In the KEM-experiment we observe a small difference in registration time for the various KEMs; in the Signature-experiment the difference is larger and SPHINCS is the faster algorithm. 

As described in the above, we mainly experimented with the effect the number of UEs has on network peformance. 
In Figure \ref{fig:numUEscaling} we can observe a distinct jump in the timings for the 95th percentile, first visible at 40 UEs. The jump is roughly 200ms, and is most likely caused by a queue batching mechanism in the system, but we have not been able to determine which mechanism specifically causes this.\footnote{There are various timers in the \texttt{ueransim} software, some of which had a default value of 200ms, but altering the \texttt{ueransim} source code to change these timers had no effect on the results.} Analyzing these results with the assumption that there is indeed an unidentified batching mechanism in mind, we see a linear increase in the 95th percentiles. We have observed more jumps (up to 400ms, 600ms, etc.) at higher UE counts and higher percentiles, but many of these are the result of the system as a whole failing. Increasing the number of UEs to 130 and higher also leads to unpredictable behaviour, such as UE requests being ignored by the core/RAN, errors and warnings being triggered in the core etc. We have therefore chosen to limit our results to 120 UEs. 

Under normal operating conditions, there is no clear difference between the conventional and post-quantum algorithms, even when looking at the results for higher percentiles. Considering outliers, however, the results of each algorithm selection diverge significantly, and the usage of post-quantum algorithms would lead to poor experience in such rare cases. This can be seen in Figure~\ref{fig:numUEscaling_100}.

\begin{table*}[ht]
    \centering
    \caption{Summary results of the PQ-TLS enabled Open5GS experiments.}
    \begin{tabular}{l|l|l|l}
                    & \multicolumn{2}{c|}{UE setup duration} &  \\
        Algorithms  & Median & 99th percentile  & SBI data rate \\ \hline 
        \texttt{(secp256r1, rsaEncryption)} & 257 ms & 505 ms & 111.0 KB/s \\ \hline
        \texttt{(p384\_bikel3, p384\_mldsa65)} & 264 ms & 669 ms & 114.8 KB/s \\ \hline
        \texttt{(p384\_bikel3, rsaEncryption)} & 260 ms & 571 ms & 112.5 KB/s \\ \hline
        \texttt{(secp256r1, p384\_mldsa65)} & 260 ms & 507 ms & 115.2 KB/s \\ \hline
    \end{tabular}
    \label{tab:pq_ogs_summary}
\end{table*}

\section{Conclusion \& Discussion}

The main conclusion of this work is a positive one: we can update the algorithms used by TLS for key establishment and authentication in the 5G Core to a post-quantum variant without a substantial impact on the usability of the network. For future versions of 5G and its successors, this is good news. 

As mentioned in the introduction, between conducting the experiments and preparing this manuscript, \cite{Scalise2024AnAA} performed a very similar experiment. This work further corroborates these earlier results and is consistent with their conclusion. 

\subsection{Future work}
Several directions for future research were out of scope for the present study and are left as future work.

\paragraph{IPsec}\label{subsubsec:5g_IPSec}
The IPsec protocol \cite{rfc4301} is used in cellular deployments to protect certain communication channels between the radio access network (RAN) and the 5G Core, including interfaces involving user-plane transport.
Versions of IPSec with support for post-quantum cryptography exist, as well as drafts standards for new protocol standards that include it.\cite{wang-ipsecme-kem-auth-ikev2-00,kampanakis-ml-kem-ikev2-09} 
Post-quantum-capable variants of IKE/IPsec have been proposed, and several Internet-Drafts describe approaches for integrating post-quantum key establishment and authentication into IKEv2~\cite{wang-ipsecme-kem-auth-ikev2-00,kampanakis-ml-kem-ikev2-09}. A natural extension of our work is to evaluate the performance and operational implications of such mechanisms within 5G deployments.

\paragraph{KEMTLS} A promising recent line of research explores instantiating TLS without conventionally generated digital signatures by replacing handshake signatures with instantiations of a (post-quantum) key-encapsulation mechanism (KEM). Especially in a post-quantum context this is promising, as the post-quantum signature algorithms from the NIST standardization project (including the ones mentioned in this work) are typically substantially more inefficient than their KEM counterparts, both in terms of computational efficiency and message size.\cite{CCS:SchSteWig20,EPRINT:GonWig22} Aditionally, they are typically more convoluted to implement.\cite[Ch. 5]{Wiggers24}
It should also be noted that various works on the topic of KEMTLS speficially point out its usability for TLS 1.3 session resumption-like use cases \cite[\S5.2]{Wiggers24}, of which the 5G core with its known components is an example.

Despite the availability reference implementations of KEMTLS exist, deploying these in the 5G Core is a challenge as the 5G Core is largely written in \texttt{C}, including the TLS implementation it uses. Perhaps the most promising implementation of KEMTLS is based on the \texttt{rustls} project,%
\footnote{\texttt{rustls}-project: \url{https://github.com/rustls/rustls/}. Accessed at commit \texttt{bdb3036}.}
\footnote{\texttt{rustls}-fork that provides KEMTLS: \url{https://github.com/thomwiggers/rustls/}. Accessed at commit \texttt{8d3925e}.} 
and while the \texttt{rustls} community has an (ongoing) project to provide FFI-bindings for easy integration with a language such as \texttt{C},%
\footnote{\texttt{rustls-ffi}-project: \url{https://github.com/rustls/rustls-ffi}. Accessed at commit \texttt{4d1d5d8}.}
the work on \texttt{C}-bindings only properly started after the KEMTLS-fork was split off from the project.%
\footnote{As of \texttt{rustls-ffi}-commit \texttt{58e2b58} (November 20, 2023), the project is no longer compatible with the KEMTLS-fork of \texttt{rustls}. Comparing this commit with the current (2025) version of \texttt{rustls-ffi}, there have been almost 19~000 updated lines.} 
This implies getting a \texttt{C}-version of KEMTLS working would either requiring manually updating the KEMTLS-fork to be up-to-date with the most recent version of \texttt{rustls-ffi} or the other way around --- either option is infeasible within the scope of this project. 

Another implementation exists in the form of a \texttt{botan}-pull request,%
\footnote{See \url{https://github.com/neXenio/botan/pull/20}.}
however there do not seem to be client or server software libraries that use \texttt{botan} and that we could easily integrate with the \texttt{open5gs} code.%
\footnote{See e.g. \url{https://github.com/randombit/botan/issues/1323} for discussion of this issue.} 
We consider experimenting with KEMTLS in the 5G Core an interesting topic for future work. 

\section*{Acknowledgments}
The research in this work was conducted as part of the CONFIDENTIAL6G project, which has received funding from the Smart Networks and Services Joint Undertaking (SNS JU) as part of the European Union's Horizon Europe research and innovation programme under Grant Agreement No.~101096435.

\bibliographystyle{IEEEtran}
\bibliography{cryptobib/abbrev3,cryptobib/crypto,Bibliography/bibliography}

\end{document}